\documentclass[12pt]{article}
\usepackage{epsfig}
\usepackage{graphicx}
\usepackage{amsmath}
\newcommand{\Tr}{\ensuremath{\mathop{\mathrm{Tr}}}}

\newcommand{\del}{\partial}
\def\be{\begin{equation}}
\def\ee{\end{equation}}
\def\bea{\begin{eqnarray}}
\def\eea{\end{eqnarray}}
\def\la{\langle}
\def\ra{\rangle}
\def\dag{\dagger}

\DeclareSymbolFont{AMSb}{U}{msb}{m}{n}
\DeclareMathSymbol{\N}{\mathbin}{AMSb}{"4E}
\DeclareMathSymbol{\Z}{\mathbin}{AMSb}{"5A}
\DeclareMathSymbol{\R}{\mathbin}{AMSb}{"52}
\DeclareMathSymbol{\Q}{\mathbin}{AMSb}{"51}
\DeclareMathSymbol{\I}{\mathbin}{AMSb}{"49}
\DeclareMathSymbol{\C}{\mathbin}{AMSb}{"43}

\def\be{\begin{equation}}
\def\ee{\end{equation}}
\def\bea{\begin{eqnarray}}
\def\eea{\end{eqnarray}}
\def\la{\langle}
\def\ra{\rangle}
\def\dag{\dagger}

\def\T{\Theta}
\def\G{\Gamma}

\def\L{\Lambda}
\def\S{\Sigma}

\def\g{\gamma}

\def\d{\delta}
\def\e{\epsilon}
\def\m{\mu}

\def\n{\nu}
\def\s{\sigma}
\def\r{\rho}
\def\l{\lambda}
\def\t{\tau}

\def\O{\Omega}

\def\N{\nabla}

\begin{document}
\title{ Exact 1/4 BPS Loop - Chiral Primary Correlator }

\author{Gordon W. Semenoff\thanks{Work supported in part by the NSERC of Canada.} and Donovan
Young\thanks{Work supported by NSERC of Canada and a University of
British Columbia Graduate Fellowship. }\\Department of Physics and
Astronomy, University of British Columbia,\\ 6224 Agricultural Road,
Vancouver, British Columbia V6T 1Z1 Canada.}

\maketitle

\begin{center}

{\bf Abstract}
\end{center}

\narrower{\narrower{\narrower{ Correlation functions of 1/4 BPS
Wilson loops with the infinite family of 1/2 BPS chiral primary
operators are computed in $\mathcal{N}=4$ super Yang-Mills theory by
summing planar ladder diagrams. Leading loop corrections to the sum
are shown to vanish. The correlation functions are also computed in
the strong-coupling limit by examining the supergravity dual of the
loop-loop correlator.   The strong coupling result is found to agree
with the extrapolation of the planar ladders.  The result is related
to known correlators of 1/2 BPS Wilson loops and 1/2 BPS chiral
primaries by a simple re-scaling of the coupling constant, similar
to an observation of Drukker, hep-th/0605151, for the case of the
1/4 BPS loop vacuum expectation value.}}}

\vskip 2cm





Recently, the study of the properties of highly symmetric states has
provided considerable insight into the AdS/CFT correspondence.
In the case of 1/2 BPS local chiral operators and 1/2 BPS Wilson
loops of $\mathcal{N}=4$ supersymmetric Yang-Mills theory, their
correspondence with 1/2 BPS gravitons and fundamental string
world-sheets has been generalized to large operators where a
beautiful picture of giant
gravitons~\cite{McGreevy:2000cw}-\cite{Corley:2001zk},
giant Wilson loops~\cite{Drukker:2005kx}-\cite{Giombi:2006de} and
bubbling geometries~\cite{Lin:2004nb} has emerged. These relate
infinite classes of highly symmetric protected operators in
Yang-Mills theory to their dual geometries which solve IIB
supergravity.

In the case of 1/2 BPS Wilson loops, an essential component of the
bubbling loop picture is the ability to compute the loop expectation
value and correlators of the loop with chiral primary operators in
Yang-Mills theory by summing planar
diagrams~\cite{Erickson:2000af}-\cite{sz},\cite{Okuyama:2006jc}. To
point, for example, it is this sum, in the form of a matrix model
computation, which provides evidence that the giant loops are dual
to D3 and D5-branes. The matrix model is thought to coincide with
the sum of all Feynman diagrams. This depends on cancellation of loop
corrections, which has been demonstrated in leading orders, but has
not yet been proven\footnote{There could also be non-perturbative
contributions, which are plausibly suppressed in the large N
limit~\cite{Bianchi:2002gz}.}. It apparently holds for the
expectation value of the 1/2 BPS Wilson loop and the correlator of
the 1/2 BPS Wilson loop with any 1/2 BPS chiral primary operator. In
all of these cases, when extrapolated to strong coupling, the sum of
planar ladder Feynman diagrams agrees with the supergravity
computation using AdS/CFT.  This gives an infinite tower of
functions which interpolate between weak and strong coupling. In
this paper, we will examine a modest extension of the picture. We
will demonstrate similar results for the expectation value and the
correlation functions of a 1/4 BPS Wilson loop with 1/2 BPS chiral
operators.

The vacuum expectation value of the 1/4 BPS loop was studied by
Drukker in Ref.~\cite{Drukker:2006ga}.  He observed a number of
interesting features of the gauge theory computation.  One was that
the ladder diagrams had a structure similar to the 1/2 BPS circle
loop and they could be summed to obtain an expression very similar
to the case of the 1/2 BPS loop. The difference was the replacement
of the 't Hooft coupling $\lambda$ by $\lambda\cos^2 \theta_0 $
where $\theta_0$ is a parameter of the 1/4 BPS loop. He further
showed that, as occurred for the 1/2 BPS loop, the leading
corrections from diagrams with internal vertices (those diagrams
which are left out of the sum over ladders) cancel. He observed
that, in the string dual where, following the prescription given in
Ref.~\cite{Maldacena:1998im}, the expectation value of the loop is
found as the area of an extremal world-sheet bounding the loop,
there are two saddle point solutions. He showed that the strong
coupling extrapolation of the sum of diagrams on the gauge theory
side carried a vestige of these two saddle points with some of the
expected features of a saddle-point expansion.

In the following, we will study correlators of 1/4 BPS  Wilson loops
with 1/2 BPS chiral primary operators.   We find that these
correlators depend on the $SO(6)$-orientation of the chiral primary.
We identify all of the orientations where the Wilson loop and the
chiral primary share some degree of supersymmetry.  We find that the
ladder diagrams can be summed for correlators of the loop and these
operators and the result is identical to those previously found with
the 1/2 BPS Wilson loop~\cite{Okuyama:2006jc}\cite{SPZ} with a
certain rescaling of the coupling constant. We shall also study the
strong coupling limit of the same correlators using the AdS/CFT
correspondence. We identify the supergravity dual of the loop-loop
correlation function and compute it in the asymptotic limit that is
appropriate to extracting the contribution of intermediate chiral
primary operators. This yields the limit of large $N$ and large 't
Hooft coupling $\lambda$. We find that the results agree with the
extrapolation to strong coupling of the Yang-Mills computation.

The Wilson loop operator of $\mathcal{N}=4$ supersymmetric
Yang-Mills theory which is most relevant to the AdS/CFT
correspondence is~\cite{Maldacena:1998im}

\begin{equation}\label{wilsonloop}
W[C]=\frac{1}{N}\Tr {\cal P} \exp\left[\int_C
\left(iA_\alpha(x(\tau))\dot x^\alpha(\tau) +|\dot
x(\tau)|\Theta^I(\tau)\Phi_I(x(\tau))\right)d\tau\right]\,,
\end{equation}
where $A_\alpha(x)$ are the gauge fields and $\Phi_I(x)$,
$I=1,...,6$ are the scalar fields of $\mathcal{N}=4$ supersymmetric
Yang-Mills theory.  The curve $C$ is described by $x^\mu(\tau)$ and
$\Theta^I(\tau)$, with $\sum_{I=1}^6\Theta^I\Theta^I=1$, describes a
loop on the 5-sphere. This loop operator is related to the holonomy
of heavy W-bosons in the gauge theory with $SU(N+1)\to SU(N)\times
U(1)$ symmetry breaking. Its string theory dual is a source for a
fundamental open string whose world-sheet ends on the contour $C$ at
the boundary of $AdS_5\times S^5$.

When probed from a distance much larger than the extension of  $C$,
the Wilson loop operator should look like an assembly of local
operators,\footnote{It is also possible to consider the insertion of
supersymmetric operators into the Wilson loop itself. We emphasize
that is a different procedure from what we are discussing here,
where correlations of primary operators with the Wilson loop are the
objects of most interest. Also, chiral operators of the type that we
consider figure promptly in the discussion of the BMN limit as well
as some issues of
integrability~\cite{Zarembo:2002ph}\cite{Pestun:2002mr}
\cite{Miwa:2005qz}\cite{Drukker:2005cu}\cite{Miwa:2006vd}.}
\begin{equation}\label{expansion}
W[C]=\la 0|W[C]|0\ra \left(1+\sum_{\Delta_i > 0} O_{\Delta_i}
(0)~L[C]^{\Delta_i} \xi_{\Delta_i}[C]\right)
\end{equation}
where $L[C]=\int_C |\dot x(\tau)|d\tau$ is the length of $C$ and we have
assumed that $C$ is near the origin 0. The operator expansion
coefficients generally depend on the shape and orientation of $C$,
as well as the parameters of $\mathcal{N}=4$ Yang-Mills theory, the
coupling constant $g_{YM}$ and the number of colors $N$. In the
remainder of this paper, we will consider only the planar 't Hooft
large $N$ limit of Yang-Mills theory where $N\to\infty$ holding
$\lambda\equiv g_{YM}^2N$ fixed. In that limit, we can see from
(\ref{correlator}) below that $\xi_\Delta $ is the ratio of a disc
to a cylinder amplitude and therefore should be of order
${\small\frac{1}{N}}$ times a function of $\lambda$.

All operators which can be made from the gauge fields, scalars and
their derivatives can appear in the expansion in
Eq.~(\ref{expansion}). We have classified operators according to
their conformal dimensions, $\Delta_i$. In a conformal field theory,
the operators of fixed conformal dimensions can be organized into
families which contain a primary operator with smallest $\Delta$ and
an infinite tower of descendants. We will assume that primary
operators are normalized so that
\be\label{normm}\la 0|O_{\Delta}(x)\,O_{\Delta'}(0)|0\ra =
\frac{\delta_{\Delta\Delta'}}{\left( 4\pi^2 x^2\right)^{\Delta}}\ee
The operator expansion coefficient $\xi_\Delta$ for a primary
operator can be extracted from the asymptotics of the correlator
\begin{equation}\label{correlator}
\frac{ \la 0|~W[C]~
O_{\Delta}(x)~|0\ra }{\la 0|~W[C]~|0\ra }
=\frac{L[C]^\Delta}{\left(4\pi^2|x|^2\right)^{\Delta}}
\xi_\Delta+\ldots\end{equation}

For example, for the 1/2 BPS circle Wilson loop,
\begin{equation}\label{halfbpsloop}C_{\tiny 1/2}:~~x^\mu(\tau)=\left(R\cos\tau,R\sin\tau,0,0\right)~~,~~
\Theta^I=\left(1,0,...\right) \end{equation} a perturbative
expansion of the loop gives
\begin{equation}\label{weak}
W[C_{\tiny 1/2}]=\la 0|~W[C_{\tiny 1/2}]~|0\ra \left( \sum_{k=0}^\infty
(2\pi R)^k \frac{ 1}{Nk!}\frac{1}{2^{k}}:{\rm Tr}(Z(0)+\bar
Z(0))^k: + ...\right)
\end{equation}
where  $Z=\left(\Phi_1+i\Phi_2\right)$ and the
dots indicate quantum corrections as well as operators with
derivatives of $Z,\bar Z$ and containing gauge fields. For the
chiral primary operators
\begin{equation}O_J\equiv\frac{1}{\sqrt{J\lambda^J}}
:{\rm Tr}Z(0)^J:\label{chirprim}\end{equation} the weak coupling
limit of $\xi_J[C_{\tiny 1/2},\lambda]$ is the appropriate
coefficient in Eq.~(\ref{weak}),
\begin{equation}\label{pert}
\xi_J[C_{\tiny 1/2};\lambda\sim 0]=\frac{1}{N}  \frac{ 1  } {2^{J}
J! }\sqrt{ J\lambda^J }
\end{equation}
This expression should receive quantum corrections.   The sum of all
quantum corrections from planar ladder diagrams was computed in
Ref.~\cite{SPZ}
\begin{equation}\label{halfbps}
\xi_J[C_{\tiny 1/2};\lambda]=\frac{1}{N} \frac{1}{2}\sqrt{\lambda
J}~\frac{ I_J(\sqrt{\lambda})}{I_1(\sqrt{\lambda})}
\end{equation}
where $I_J(x)$ is the $J$-th modified Bessel function of the first
kind.  In the expression (\ref{halfbps}), as it must, the leading
term in a small $\lambda$ expansion agrees with (\ref{pert}).  The
leading order planar diagrams which are left out of the sum over
ladders was also computed in Ref.~\cite{SPZ} and were shown to
cancel identically. It was then tempting to conjecture that these
corrections vanish to all orders. To support this conjecture, the
extrapolation of Eq.~(\ref{halfbps}) to large $\lambda$ can be
compared with the result of a computation of the same coefficients
using the AdS/CFT correspondence, originally done in
Ref.~\cite{Berenstein:1998ij},
\begin{equation}\label{strong}
\xi_J[C_{\tiny 1/2};\lambda\sim\infty]= \frac{1}{N} \frac{1}{2}~
\sqrt{\lambda J}
\end{equation}
This coincides with the large $\lambda$ limit of the expression in
Eq.~(\ref{halfbps}).  The coefficients $\xi_J[C_{\tiny
1/2},\lambda]$ in (\ref{halfbps}), together with  the result of
Ref.~\cite{Erickson:2000af} \be\label{expval} \la W[C_{\tiny 1/2}]
\ra = \frac{2}{\sqrt{\l}}\,I_1(\sqrt{\l}), \ee yield an infinite
family of interpolating functions which match both the strong and
weak coupling limits computed in string and gauge theory,
respectively.

In the present paper, we will examine the 1/4 BPS loop which has the
trajectory
\begin{equation}\label{quarterloop}C_{\tiny 1/4}:
x^\mu(\tau)=R\left(\cos \tau,\sin \tau,0,0\right),~
\Theta^I(\tau)=\left( \sin\theta_0\cos\tau, \sin\theta_0\sin\tau,
\cos\theta_0,0,0,0\right)
\end{equation}
 The main difference from the 1/2 BPS loop
is that $\Theta^I(\tau)$ moves in a circle on an $S^2\subset S^5$,
rather than sitting at a point. Putting $\theta_0$ to zero recovers
the 1/2 BPS loop in (\ref{halfbpsloop}). The special case of this
1/4 BPS loop with $\theta_0=\pi/2$ was originally discussed by
Zarembo~\cite{Zarembo:2002an}.

To understand the supersymmetries of the loop with trajectory
(\ref{quarterloop}) we recall that the supersymmetry transformation
of $\mathcal{N}=4$ Yang-Mills theory is generated by the spinor
\begin{equation} \epsilon(x)=\epsilon_0+\gamma_\mu x^\mu \epsilon_1
\end{equation} Here, we have to consider both Poincare
supersymmetries, with constant spinor $\epsilon_0$ and conformal
supersymmetries, with constant spinor $\epsilon_1$. In order to be
supersymmetries of the 1/4 BPS Wilson loop, it is straightforward to
see that they have to satisfy the equations~\cite{Drukker:2006ga}

\bea\label{Wsusy1} \sin\theta_0\left( \g^1 \G^2 + \g^2 \G^1 \right)
\e_0 = 0
\qquad \sin\theta_0\left( \g^1 \G^2 + \g^2 \G^1 \right) \e_1 = 0\\
\label{Wsusy2} \cos \theta_0 \, \e_0 = R\left(-i\g^1 + \sin \theta_0
\, \G^2 \right) \, \G^3\g^2\,\e_1 \eea where the ten dimensional
gamma matrices are $(\gamma^i,\Gamma^I)$ with $i=1,...,4$ and
$I=1,...,6$. Let us count the supersymmetries.  Each of the spinors
$\epsilon_0$ and $\epsilon_1$ has 16 components.  The conditions in
(\ref{Wsusy1}) are half-rank and reduce the number of each of the
spinors by half. Then (\ref{Wsusy2}) relates the remaining
components of $\epsilon_1$ to those of $\epsilon_0$ in a way which
is compatible with (\ref{Wsusy1}). The remaining independent
components are eight -- half of the original 16 components of
$\epsilon_0$.  This is 1/4 of the original 32 components of
$\epsilon_0$ and $\epsilon_1$.

We will consider a chiral operator which has an arbitrary $SO(6)$
orientation, beginning with
$$
{\rm Tr}\left( u\cdot\Phi(0)\right)^J
$$
where $u$ is a complex 6-vector, satisfying the constraint that
$u^2=0$.   Being a scalar operator, conformal supersymmetries are
automatic.  This operator has some Poincare supersymmetry if there
exist some non-zero constant spinors $\epsilon_0$ which solve the
equation
\begin{equation}\label{susy}
u\cdot\Gamma \epsilon_0=0
\end{equation}
There are solutions  only when $\left(u\cdot\Gamma\right)^2=u^2=0$
which, as we have assumed, is the case. Then $u\cdot\Gamma$ is
half-rank and there are exactly eight independent non-zero solutions
of Eq.~(\ref{susy}).

Now we can ask the question as to whether the eight independent
$\epsilon_0$ which solve (\ref{susy}) have anything in common with
the eight solutions of (\ref{Wsusy1}) and (\ref{Wsusy2}), i.e. are
there spinors which solve both of them?

Before we answer this question, let us backtrack to the case of the
1/2 BPS loop geometry (\ref{halfbps}).  There Eq.~(\ref{Wsusy1}) is
absent and the spinors must solve   (\ref{Wsusy2}) with
$\theta_0=0$. This simply relates $\epsilon_1$ to $\epsilon_0$,
eliminating half of the possible spinors.  There are 16 independent
solutions of this equation -- it is 1/2 BPS.  Now, consider a chiral
primary operator.  Without loss of generality, we can consider the
operator ${\rm Tr}\left(\Phi_1+i\Phi_2\right)^J$. It is
supersymmetric if $\epsilon_0$ satisfies the equation $$\left(
\Gamma^1+i\Gamma^2\right)\epsilon_0=0$$  The matrix
$\Gamma_1+i\Gamma_2$ has half-rank, so this requirement eliminates
half of the supersymmetries generated by $\epsilon_0$.  This leaves
eight supersymmetries which commute with both the 1/2 BPS Wilson
loop and the 1/2-BPS chiral primary operator. This high degree of
residual joint supersymmetry is thought to be responsible for the
fact that, apparently, only ladder diagrams contribute to the
asymptotic limit of their correlator.

Returning to the 1/4 BPS loop and chiral primary with general
orientation, it is easy to see that there is a simultaneous solution
of (\ref{Wsusy1}), (\ref{Wsusy2}) and (\ref{susy}) only when one of
the following holds:
\begin{itemize}\item{}$u_1=u_2=0$.  We can always do an $SO(6)$ rotation
which commutes with the loop operator and sets
$(u_4,u_5,u_6)\to(u_4,0,0)$. Then,  there will be simultaneous
solutions of (\ref{Wsusy1}), (\ref{Wsusy2}) and (\ref{susy}) only
when $u_3= iu_4$ or when $u_3=-iu_4$. In both of these cases, there
are four solutions, corresponding to 1/8 supersymmetry in common
between the chiral primary and the Wilson loop.  Up to a constant,
the chiral primary operator is ${\rm Tr}\left( \Phi_3 +
i\Phi_4\right)^J$ or the complex conjugate ${\rm Tr}\left( \Phi_3 -
i\Phi_4\right)^J$.
\item{}$u_3=u_4=0$. There is a solution when $u_1=\pm iu_2$ and there
is also 1/8 supersymmetry. The chiral primary is ${\rm Tr}\left(
\Phi_1+ i\Phi_2\right)^J$ or its complex conjugate. In this case, we
show in Appendix \ref{app3} that the coefficient $\xi_J$ which is
extracted from the long range part of the correlator of this
operator and the loop vanishes due to R-symmetry. Thus, for all
$J\ra 0$, the coefficients of ${\rm Tr}\left( \Phi_1+ i\Phi_2\right)^J$
or ${\rm Tr}\left( \Phi_1- i\Phi_2\right)^J$ in the operator
expansion of the 1/4 BPS loop are zero.
\item{}$u_1=\pm iu_2$. There are two non-zero solutions when $u_3=
iu_4$ or when $u_3=-iu_4$.  This corresponds to 1/16 supersymmetry.
There are essentially four operators,
$$ {\rm Tr}\left( \chi\left(\Phi_1+i\Phi_2\right)+
\left(\Phi_3+i\Phi_4\right)\right)^J $$ plus others with
substitutions of $\Phi_1-i\Phi_2$ or $\Phi_3-i\Phi_4$. In this case
too, because of R-symmetry the contribution with any non-zero power
of $\left( \Phi_1\pm i\Phi_2\right)$ will be zero.  The coefficient
$\xi_J[C_{\tiny 1/4}]$ for these operators is therefore the same as
those for the operator ${\rm Tr}\left(\Phi_3\pm i\Phi_4\right)^J $.
\end{itemize}
Thus we see that the interesting quantity where there is some degree
of supersymmetry common to both the loop operator and the primary is
\begin{equation}\label{quartbps}
\xi_J[C_{\tiny 1/4}]=\lim_{|x|\to\infty} \left(
\frac{4\pi^2|x|^2}{2\pi R}\right)^J
\frac{1}{\sqrt{J\lambda^J}}\frac{\la 0|~ W[C_{\tiny 1/4}] ~{\rm
Tr}\left( 
\Phi_3(x)+i\Phi_4(x)\right)^J~|0\ra }{\la 0|~ W[C_{\tiny
1/4}]~|0\ra }
\end{equation}
 It is these partially
supersymmetric configurations which we expect to have some level of
protection from quantum corrections. Indeed, we shall find evidence
for this.    All other possibilities either vanish, are equivalent
to (\ref{quartbps}) or have no supersymmetry at all. The cases with
no supersymmetry at all are apparently not protected.

\begin{figure}
\label{leading} \centerline{
   \includegraphics[bb=419 279 190 508,height=2in]{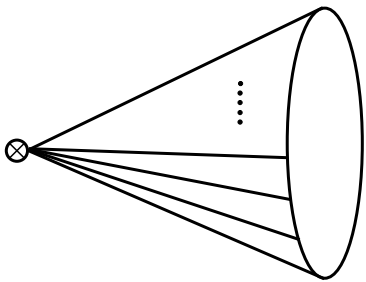}
} \caption{The leading planar contribution to $\la
   W[C_{\tiny 1/4}]~{\rm Tr}(\Phi_3+i\Phi_4)^J \ra$.
   There are $J$ lines connecting the chiral
   primary on the left with the circular Wilson loop on the right. }
\end{figure}

We will present arguments that the sum of planar ladder diagrams
contributing to the correlation function in (\ref{quartbps}) gives a
contribution which differs from the one for the 1/2 BPS loop quoted
in Eq.~(\ref{halfbps}) by the simple replacement
$\lambda\to\lambda\cos^2\theta_0$, so that the total result is
\begin{equation}\label{quarterbps}
\xi_J[C_{\tiny 1/4}]=\frac{1}{N}\frac{1}{2}
~\sqrt{\lambda\cos^2\theta_0 J}~\frac{
I_J(\sqrt{\lambda\cos^2\theta_0})}{I_1(\sqrt{\lambda\cos^2\theta_0})}
\end{equation}
To find this result using Feynman diagrams, we begin with the lowest
order diagrams, depicted in Fig.1.  There, each occurrence of the
scalar $\Phi_3$ in the composite operator contracts with a scalar
$\Phi_3$ in the Wilson loop. We consider only the planar diagrams.
Each scalar $\Phi_3$ from the Wilson loop carries a factor of
$\cos\theta_0$, leading to an overall factor of $(\cos\theta_0)^J$.
We are taking the convention for Feynman rules where each line in
the Feynman diagram results in a factor of $\lambda$, totaling
$\lambda^J$ for the diagram in Fig.1. With this convention, the
chiral primary operator has normalization $\lambda^{-J/2}$ (see
(\ref{chirprim})). The net result is a factor of $\lambda^{J/2}$
which combines with the  $(\cos\theta_0)^J$ to give a coupling
constant dependence in the form $(\lambda\cos^2\theta_0)^{J/2}$.
This is identical to what one would have obtained by taking the same
diagram for the 1/2 BPS loop and simply replacing $\lambda$ by
$\lambda\cos^2\theta_0$.

To compute the next orders, we must decorate the diagram in Fig.1
with propagators.  The simplest are ladder diagrams, see Fig.2, which go
between two points on the periphery of the loop. They are
described by summing the contribution of the vector and the scalar
field. In the Feynman gauge, the sum of scalar and vector
propagators connecting two points on arcs of the same circle is a
constant:
\begin{eqnarray}
 |\dot x(\sigma)|\Theta^I(\sigma)\,\la\Phi_I(x(\sigma))\Phi_J(x(\tau))\ra\,|\dot
x(\tau)|\Theta^J(\tau)-\dot x^\alpha(\sigma)\,\la
A_\alpha(x(\tau))A_\beta(x(\tau))\ra\, \dot x^\beta(\tau) \nonumber
\\=\frac{ |\dot x(\sigma)||\dot
x(\tau)|\Theta(\sigma)\cdot\Theta(\tau)-\dot x(\sigma)\cdot\dot
x(\tau) }{ 4\pi^2\left( x(\sigma)-x(\tau)\right)^2 }
=\frac{R^2}{8\pi^2}\cos^2\theta_0 \nonumber \end{eqnarray} This
is what makes ladder diagrams easy to sum. We note that this
propagator is accompanied by a factor of $\lambda$, so the total
$\lambda$ and $\theta_0$-dependence again comes in the combination
$\lambda\cos^2\theta_0$. Further, the only difference from the
analogous quantity for the 1/2 BPS loop is the factor
$\cos^2\theta_0$. Thus we see that the sum of ladders for this 1/4
BPS loop will be identical to that for the 1/2 BPS loop with the
replacement $\lambda\to \lambda \cos^2\theta_0$.

\begin{figure}
 \centerline{
\includegraphics[height=1.2in,angle=180]{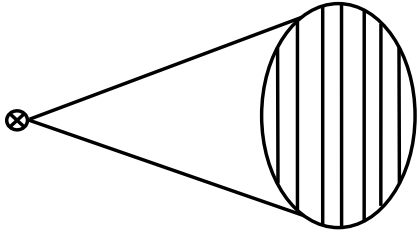}
} \caption{A ladder diagram of $\la W[C_{\tiny
1/4}]~{\rm Tr}(\Phi_3+i\Phi_4)^J \ra$. The ``rungs'' represent the combined gauge
   field and scalar propagator. For clarity, $J$ has been set to 2.}
\end{figure}

\begin{figure}
\label{fig2} \centerline{
   \includegraphics[height=3in]{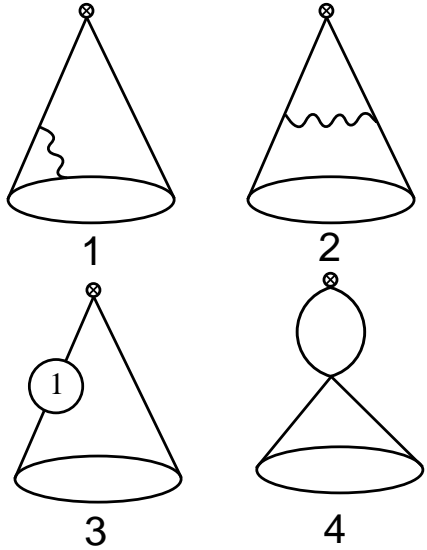}
} \caption{The one-loop radiative corrections to $\la W[C_{\tiny
1/4}]~{\rm Tr}(\Phi_3+i\Phi_4)^J \ra$. Only an adjacent pair of the
$J$ scalar lines are shown. }
\end{figure}

Finally, there are the diagrams that have not yet been included so
far. The conjecture is that they vanish.  The leading order are
depicted in Fig.3. By a simple generalization of the argument
obtained in Ref.~\cite{SPZ} and explained in more detail in
Ref.~\cite{Pestun:2002mr}, they can be shown to cancel identically.
Assuming that this cancellation occurs to higher orders as well, the
result for the summation of all planar Feynman diagrams is
summarized in the formula (\ref{quarterbps}).

We now turn to the string theory dual of the correlator of the 1/4
BPS Wilson loop and the chiral primary operator.  This will give a
strong coupling planar limit of the operator expansion
coefficients. It is most efficient to extract the operator
expansion coefficient from the asymptotic form of the connected
correlator of two Wilson loops, where the contributions of chiral
primary intermediate states can be easily identified. This was
used to compute the same quantity for a 1/2 BPS loop in
Ref.~\cite{Berenstein:1998ij}. The string theory dual of the
Wilson loop operator is a fundamental string worldsheet which has
as boundary the contour $C$ and which itself sits at the boundary of
the space $AdS_5\times S^5$~\cite{Maldacena:1998im}. The coupling
constant of the string sigma model is
$\alpha'/\mathcal{R}^2=1/\sqrt{\lambda}$ where $\mathcal {R}$ is
the radius of curvature of $AdS_5\times S^5$ and we have used its
relation with the 't Hooft coupling
$\mathcal{R}^4/{\alpha'}^2=\lambda$. In the limit of large
$\lambda$, the worldsheet sigma model is weakly coupled and can be
solved semi-classically.  The leading order is classical, it
simply finds an extremal surface with boundary $C$ and which is
compatible with other boundary conditions.

The connected loop-loop correlator has an extremal surface whose
boundary is the two loops.  When the loops have large separation,
this surface degenerates to two disc geometry worldsheets whose
boundaries are each loop with an infinitesimal tube connecting
them, see figure Fig.4. In the limit of large separation, this tube is
described by
the propagator of the lightest gravity modes, which at large
$\lambda$ are 1/2 BPS supergravitons, the string theory duals of
the chiral primary operators. The connection between the graviton
propagator and the worldsheet is through a vertex operator which
must be identified and the connection point with the vertex
operator must be integrated over the worldsheet. The resulting
amplitude is proportional to the square of the desired operator
expansion coefficient.

To begin, the first step is to identify the minimal surface in
$AdS_5\times S^5$ whose boundary is the 1/4 BPS contour $C_{\tiny
1/4}$. This was done in Ref.~\cite{Drukker:2006ga}. We will
summarize it here in more convenient coordinates.  We take the
metric of $AdS_5 \times S^5$

\bea\label{metric} ds^2 &=& \sqrt{\lambda}\left( \frac{ dy^2 + d
r_1^2 + r_1^2 d\phi_1^2 + d r_2^2 + r_2^2 d \phi_2^2  }{y^2} \right.
\nonumber \\  &+& \left.
  d\theta^2 + \sin^2 \theta d\phi^2 + \cos^2
\theta \left( d\r^2 + \sin^2\r\, d\hat\phi^2 + \cos^2 \r\, d \tilde
\phi^2 \right)  \right) \eea  The string world-sheet is then
embedded as follows,
\begin{eqnarray}y = R \tanh \s \qquad r_1 =
\frac{R}{\cosh \s} \qquad \phi_1 = \t \qquad r_2=0 \qquad \phi_2={\rm const.}
\nonumber \\
 \sin \theta = \frac{1}{\cosh(\s_0 \pm \s)} \qquad \phi = \tau
 \qquad \rho=\frac{\pi}{2} \qquad \hat\phi=0 \qquad \tilde\phi={\rm const}.
\label{embed} \end{eqnarray}  where $\s \in [0,\infty]$ and $\t\in
[0, 2\pi]$ are the world-sheet coordinates. The  contour $C_{\tiny
1/4}$ is the boundary of the worldsheet at $\s=0$, which in turn
sits at $y=0$, the boundary of $AdS_5\times S^5$. The parameter
$\cos\theta_0=\frac{1}{\cosh\sigma_0}$.  The choice of $\pm$ sign in
the embedding of $\theta$ arises because there are two saddle points
in the classical action corresponding to wrapping the north or south
pole of the $S^5$. Of course the sign should be chosen to minimize
the classical action, which corresponds to choosing +. The other
saddle point is unstable, and the string world-sheet will slip-off
the unstable pole.


\begin{figure}
\centerline{
   \includegraphics[height=1.75in]{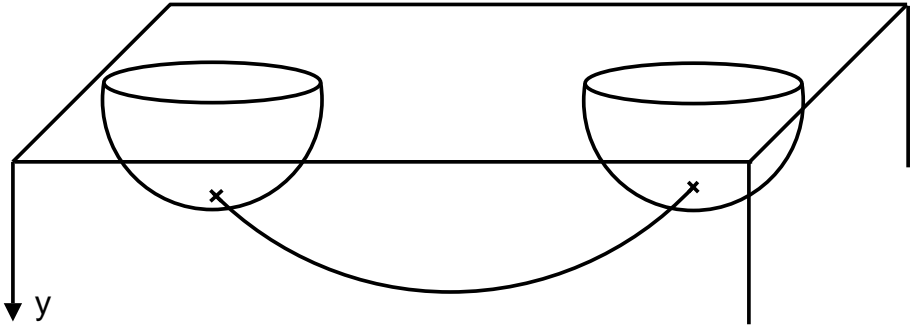}
} \caption{The string worldsheets of two widely separated Wilson loops
  exchange a supergravity mode dual to a chiral primary operator.}
\end{figure}

The supergravity modes that we are interested in are fluctuations of
the RR 5-form as well as the spacetime metric. They are by now very
well known, and details can be found in
Refs.~\cite{Berenstein:1998ij}\cite{Kim:1985ez}\cite{Lee:1998bx}\cite{Semenoff:2004qr}\cite{Giombi:2006de}.
The fluctuations are \bea\label{fluct} \d g_{\alpha \beta} &=&
\left[-\frac{6\,J}{5}\,g_{\alpha \beta} + \frac{4}{J+1} \,
  D_{(\alpha} D_{\beta)} \right] \,s^J(X)\,Y_J(\O),\cr
\d g_{IK } &=& 2\,k\,g_{IK } \,s^J(X)\,Y_J(\O), \eea where
$\alpha,\beta$ are $AdS_5$ and $I,K$ are $S^5$ indices. The symbol
$X$ indicates coordinates on $AdS^5$ and $\O$ coordinates on the
$S^5$. The $D_{(\alpha} D_{\beta)}$ represents the traceless
symmetric double covariant derivative. The $Y_J(\O)$ are the
spherical harmonics on the five-sphere, while $s^J(X)$ have
arbitrary profile and represent a scalar field propagating on
$AdS_5$ space with mass squared $=J(J-4)$, where $J$ labels the
representation of $SO(6)$ and must be an integer greater than or
equal to 2. (This is the representation of $SO(6)$ which contains
the chiral primary operators that we are interested in.)

The supergravity field dual to the operator
$\Tr\left(u\cdot\Phi\right)^J$ is
obtained by choosing the combination of spherical harmonics with the
same quantum numbers and evaluating them on the worldsheet using
(\ref{embed}) (see appendix \ref{app2}) so that, \be
Y_J(\theta,\phi)= {\cal N}_J(u) \biggl[ u_1\sin \theta \,\cos\phi + u_2 \sin
\theta \,\sin\phi + u_3 \cos \theta \biggr]^J  \ee  The worldsheets
will be connected by the propagator for the scalar supergravity mode
$s^J(X)$.  The asymptotic form of this propagator for large
separation $x$ is  \be\label{prop} P(X,\bar X ) = \la s^J(X)\,
s^J(\bar X) \ra \simeq \L_J \, \left( \frac{1}{x} \right)^{2J} \,
y^J\, \bar y^J \ee where $\L_J = 2^J (J+1)^2 /(16\, N^2 J)$.  The
barred quantities are coordinates on the second Wilson loop
worldsheet. Then, in the large $\lambda$ limit, the Wilson loop
correlator is \be \frac{\la 0|~ W[C_{\tiny 1/4},x]~ W^*[C_{\tiny
1/4},0]~|0\ra  }{ \left|\la 0|~
  W[C_{\tiny 1/4}]~|0\ra \right|^2 }   =
\int_{\S} \int_{\bar \S} \del_a X^M \del^a X^N \,\d g_{MN}\,
P(X,\bar X) \, \d \bar g_{\bar M \bar N} \,\del_{\bar a} X^{\bar M}
\del^{\bar a} X^{\bar N}, \ee where $M,N=1,...,10$ and the $\delta g_{MN}$ are
given in (\ref{fluct}), except now we have removed the fluctuating
parts, $s^J(X)$ and replaced them by the propagator $P$.   The
pullback of the fluctuations (\ref{fluct}) to the worldsheet are
found in appendix \ref{app1}. Using them we have, \bea\label{integr}
 \frac{\la 0|~ W[C_{\tiny 1/4},x]~ W^*[C_{\tiny 1/4},0]~|0\ra  }{
\left|\la 0|~
  W[C_{\tiny 1/4}]~|0\ra \right|^2 }
= \frac{\L_J}{x^{2J}}\frac{\lambda}{16 \pi^2 } \Biggl[ 2J
 \int
d\s d\t y'^2 y^{J-2}   Y_J(\theta,\phi)- ~~~~~~~~~~\cr ~~~~~~~ -  2J
 \int
d\s d\t (r_1'^2+r_1^2) y^{J-2}   Y_J(\theta,\phi) +  2J \int d\s d\t
(\theta'^2+\sin^2 \theta) y^J   Y_J(\theta,\phi)  \Biggr]^2 \eea
Each of the terms inside the square on the right-hand-side of the
above expression has a common factor of

\bea\label{str} \int_0^{2\pi}d\tau \,Y_J(\theta,\phi)= 
{\cal N}_J(u) \int_0^{2\pi}
d\t  \Bigl[ u_1\sin\theta \cos\t +u_2\sin\theta \sin\t +
u_3\cos\theta \Bigr]^J
\eea From this expression we see that, consistent with our
expectations using R-symmetry on the gauge theory side, for the at
least 1/16 supersymmetric combination of loop and primary when
$u_2=\pm iu_1$, the dependence on $u_1$ and $u_2$ integrates to
zero.  If these parameters are chosen more arbitrarily, so that
there is no supersymmetry at all, the loop depends on them. In that
case the contributions proportional to powers of $u_1$ and $u_2$ in
the final result for the operator expansion coefficients do not
follow the rule that they are related to the 1/2 BPS loop ones by
the replacement of $\lambda$ by $\lambda\cos^2\theta_0$.  We
attribute this to absence of supersymmetry.  From here, we will
proceed with the supersymmetric case only by putting $u_1=u_2=0$ and
$u_3=1$.

We will now compute the integrals in (\ref{integr}) with this
assumption. We note that the embedding (\ref{embed}) has some nice
properties. For instance $y'^2 + r_1'^2 = r_1^2 = y'$ and also
$\sin^2 \theta = \theta'^2$. Using these, we can express the
integrals in (\ref{integr}) as follows, \bea \frac{2^{-J/2}}{R^J}
\int d\s y'^2 y^{\small J-2}   \cos^J \theta = 2^{-J/2}\int_0^\infty
d\s \frac{(\tanh \s)^{J-2}}{\cosh^4 \s}  \tanh^J (\s_0 \pm \s) \cr =
2^{-J/2}\int_0^1 dz (1-z^2)z^{J-2}\left(\frac{\pm z + \cos
  \theta_0}{1 \pm z \cos \theta_0}\right)^J
\\  \frac{2^{-J/2}}{R^J} \int d\s (r_1'^2+r_1^2) y^{\tiny J-2}   \cos^J \theta
 =2^{-J/2} \int_0^1 dz (1+z^2)z^{J-2}\left(\frac{\pm z + \cos
  \theta_0}{1 \pm z \cos \theta_0}\right)^J
\\ \frac{2^{-J/2}}{R^J} \int d\s (\theta'^2+\sin^2 \theta) y^J  \cos^J
\theta  = -2^{1-J/2}\int_{\mp \cos \theta_0}^{-1} dz
 \left(\frac{ \pm z + \cos
  \theta_0}{1 \pm z \cos \theta_0}\right)^J  z^J
\eea
  Putting everything together,

\bea\label{final} &&  \frac{\la 0|~ W[C_{\tiny 1/4},x]~ W^*[C_{\tiny
1/4},0]~|0\ra  }{ \left|\la 0|~
  W[C_{\tiny 1/4}]~|0\ra \right|^2 }
 = \cr
&&~~=16\,J^2\,
 \frac{\L_J}{2^J}\,\left(\frac{R}{x}\right)^{2J}\,\frac{\lambda}{4
}\Biggl[ \left\{ \int_{-1}^{\mp \cos \theta_0} dz - \int_0^1 dz
\right\} \left( \frac{\pm z + \cos \theta_0}{1\pm z\cos \theta_0}
\right)^J \, z^J\Biggl]^2\cr &&~~= 16\,J^2\,
\frac{\L_J}{2^J}\,\left(\frac{R}{x}\right)^{2J}\,\frac{\lambda}{4 }\left[
\frac{-(\pm)^{J+1} \cos \theta_0}{J+1} \right]^2 =
\frac{1}{4N^2}\,J\,\lambda\cos^2
\theta_0\left(\frac{R}{x}\right)^{2J}, \eea  which is just the
result for the 1/2 BPS circle \cite{Berenstein:1998ij} with $\l
\rightarrow \l \cos^2 \theta_0$. Using the prescription
\cite{Berenstein:1998ij} to obtain from the loop-to-loop correlator
the overlap with the chiral primary in question, we find
$\xi_J[C_{\tiny 1/4}]=\sqrt{J\lambda\cos^2\theta_0}/2N$.
  This is identical to the large
$\lambda$ limit of Eq.~(\ref{quarterbps}). We have thus confirmed
that the sum of planar ladder diagrams agrees with the prediction of
AdS/CFT in the strong coupling limit. The emergence of this
structure on the supergravity side of the duality is non-trivial.
The integrations over the $AdS_5$ and $S^5$ portions of the string
worldsheet conspire in a complicated way in (\ref{final}) to give
the $\l \rightarrow \cos^2\theta_0\,\l$ result.

It is instructive to consider  this calculation where both saddle
points of the classical action are kept in the path integral, as is
discussed in \cite{Drukker:2006ga}. There it was noted that the
semi-classical result for the expectation value of the Wilson loop
is a sum of two terms; one proportional to $\exp(\sqrt{\l'})$ and
the other to $\exp(-\sqrt{\l'})$, where $\l'=\cos^2 \theta_0\, \l$.
This was mirrored in the asymptotic expansion \cite{GR} of the
modified Bessel function of (\ref{expval}),

\be
\begin{split}
&I_1(\sqrt{\l'}) =\\ &\frac{e^{\sqrt{\l'}}}{\sqrt{2\pi\sqrt{\l'}}}\,
\sum_{k=0}^{\infty} \left( \frac{-1}{2 \sqrt{\l'}} \right)^k\,
\frac{ \G(3/2+k) }{ k!\,\G(3/2-k) } \pm
i\, \frac{e^{-\sqrt{\l'}}}{\sqrt{2\pi\sqrt{\l'}}}\,
\sum_{k=0}^{\infty} \left( \frac{1}{2 \sqrt{\l'}} \right)^k\,
\frac{ \G(3/2+k) }{ k!\,\G(3/2-k) },
\end{split}
\ee  where the sign of the $i$ is ambiguous due to the {\it Stokes'
Phenomenon} \cite{watson}. The factor of $i$ was associated with the
fluctuation determinant of the three tachyonic modes associated with
the worldsheet slipping off the unstable pole of the five-sphere.

Due to the sign structure found in (\ref{final})
before squaring, the analogous structure for the connected correlator of the
primary with the loop
is a sum of a term proportional to $\exp(\sqrt{\l'})$ and of another
proportional to  $(-1)^{J+1}\,\exp(-\sqrt{\l'})$. The sum of these two
terms should then be normalized by the expectation value of the Wilson
loop. If we employ the
asymptotic expansions of the modified Bessel
functions in (\ref{halfbps}), we have

\be\begin{split}
&\frac{I_J(\sqrt{\l'})}{I_1(\sqrt{\l'})} =\\
&\frac{ e^{\sqrt{\l'}}\,
\sum_{k=0}^{\infty} \left( \frac{-1}{2 \sqrt{\l'}} \right)^k\,
\frac{ \G(J+k+1/2) }{ k!\,\G(J-k+1/2) } \mp
i\, (-1)^J\,e^{-\sqrt{\l'}}\,
\sum_{k=0}^{\infty} \left( \frac{1}{2 \sqrt{\l'}} \right)^k\,
\frac{ \G(J+k+1/2) }{ k!\,\G(J-k+1/2) } }
{ e^{\sqrt{\l'}}\,
\sum_{k=0}^{\infty} \left( \frac{-1}{2 \sqrt{\l'}} \right)^k\,
\frac{ \G(3/2+k) }{ k!\,\G(3/2-k) } \pm
i\, e^{-\sqrt{\l'}}\,
\sum_{k=0}^{\infty} \left( \frac{1}{2 \sqrt{\l'}} \right)^k\,
\frac{ \G(3/2+k) }{ k!\,\G(3/2-k) } }.
\end{split}
\ee This clearly reflects the presence of two saddle points in the
functional integrals in both the numerator and denominator.

We also note that the chiral primary has zero overlap with the
supersymmetric Wilson loop (i.e. $W_{\theta_0=\pi/2}$). This is
expected, since two such Wilson loops should not interact with each
other by supersymmetry.

There has been extensive work of late concerning Wilson loops whose
$SU(N)$ representations are of higher rank \cite{Drukker:2005kx}-\cite{Rodriguez-Gomez:2006zz},
\cite{Lunin:2006xr}-\cite{Hartnoll:2006ib}. They have been associated
with D-brane solutions analogous to giant gravitons. Explicit
solutions are available for the 1/2 BPS loop, and results have been matched
to matrix model calculations. It would be very
interesting to solve the DBI equations of motion corresponding to the
1/4 BPS loop, and to repeat the calculations done here for that
solution, as has been recently done for the 1/2 BPS case \cite{Giombi:2006de}.


\appendix

\section{Metric Fluctuations}
\label{app1}

Given (\ref{fluct}) and (\ref{metric}), we must construct the traceless
symmetric double covariant derivative,

\be
D_{(\m} D_{\n)} \equiv \frac{1}{2} \left( D_\m D_\n + D_\n D_\m \right) -
\frac{1}{5} g_{\m \n} \, g^{\r \s} D_{\r \s}.
\ee

\noindent The action of $D_\m D_\n$ on a scalar field $\phi$
is,

\be
D_\m D_\n \phi = \del_\m \del_\n \phi - \G^\l_{\m \n} \del_{\l} \phi.
\ee

\noindent The Christoffel symbols for the $AdS$ geometry (\ref{metric}) are,

\bea
\G^{r_i}_{\phi_i \phi_i} = -r_i \qquad \G^y_{\phi_i \phi_i} =
\frac{r_i^2}{y} \qquad &&
\G^{\phi_i}_{\phi_i r_i} = \frac{1}{r_i} \qquad \G^{\phi_i}_{\phi_i y}
= -\frac{1}{y} \cr
\G^{y}_{r_i r_i} = \frac{1}{y} \qquad \G^{r_i}_{y r_i} &=&
-\frac{1}{y} \qquad \G^{y}_{y y} = -\frac{1}{y}
\eea

\noindent where $i=1,2$. The trace of $D_\m D_\n \,\phi$ is given by,

\be
g^{\m \n} D_\m D_\n = \sum_{i=1}^2 \left( y^2 \del_y^2 + y^2 \del_{r_i}^2 +
\frac{y^2}{r_i^2} \del_{\phi_i}^2 -3y \del_y + \frac{y^2}{r_i} \del_{r_i}
\right)\,\phi
\ee

\noindent Because of (\ref{prop}), we only keep those terms of $D_{(\m} D_{\n)}$
which contain derivatives in $y$. These are,

\be
D_{(y} D_{y)} = \frac{4}{5} \del_y^2  + \frac{8}{5 y} \del_y, \qquad
D_{(r_1} D_{r_1)} = \frac{1}{r_1^2} D_{(\phi_1} D_{\phi_1)} =  -\frac{1}{5} \del_y^2
-\frac{2}{5y} \del_y.
\ee

\noindent We now note that since the derivatives will be acting on
$y^J$ from the propagator, we may replace $\del_y^2 \rightarrow
J(J-1)/y^2$ and $y^{-1}\del_y \rightarrow J/y^2$.  Therefore the
metric fluctuations may be expressed as follows,

\bea
\d g_{y y} &=&
\left[ -\frac{6J}{5} + \frac{4}{J+1} \left( \frac{4}{5} J(J-1) +
  \frac{8}{5}J\right) \right] \frac{L^2}{y^2} = 2J \frac{L^2}{y^2}\cr
\d g_{r_1 r_1} &=& \frac{1}{r_1^2} \d g_{\phi_1 \phi_1} =
\left[ -\frac{6J}{5} -\frac{4}{J+1} \left( \frac{1}{5} J(J-1) +
  \frac{2}{5}J\right) \right] \frac{L^2}{y^2} = -2J \frac{L^2}{y^2}.
\eea

\section{Spherical Harmonics}
\label{app2}

The five-sphere is embedded in $\R^6$ in the following manner,

 \bea &&x^1 = \sin \theta \cos \phi \qquad \qquad x^2
= \sin \theta \sin \phi \cr &&x^3 = \cos \theta \sin \r \cos \hat
\phi \qquad x^4 = \cos \theta \sin \r \sin \hat \phi \cr &&x^5 =
\cos \theta \cos \r \cos \tilde \phi \qquad x^6 = \cos \theta \cos
\r \sin \tilde \phi, \eea and has the metric

\be ds^2_{S^5} = d\theta^2 + \sin^2 \theta \,d\phi^2 + \cos^2 \theta
\, \left( d\r^2 + \sin^2\r\, d\hat\phi^2 + \cos^2 \r\, d \tilde
\phi^2 \right). \ee The embedding (\ref{embed}) takes $\r = \pi/2,
\hat \phi =0$, or $x^4 =x^5=x^6=0$. Note that $\r \in [0,\pi/2]$ while
$\theta \in [0,\pi]$. A general chiral primary normalized as in
(\ref{normm}) may be written as,

\be
\frac{2^{J/2}}{\sqrt{\l^J J}} C^{I_1 \ldots I_J} \Tr 
\Phi_{I_1} \ldots \Phi_{I_J}
\ee

\noindent where $C^{I_1 \ldots I_J}$ is traceless symmteric and
$C^{I_1 \ldots I_J}{C^*}^{I_1 \ldots I_J} = 1$. The corresponding
spherical harmonic
is given by $Y_J(\theta,\phi) = C^{I_1 \ldots I_J} x^{I_1} \ldots
x^{I_J}$. A properly normalized (i.e. (\ref{normm})) operator built on 
$\Tr(u \cdot \Phi)^J$ will then correspond to

\be
Y_J(\theta,\phi) ={\cal N}_J(u) \biggl[
u_1 \sin \theta \,\cos\phi +
u_2 \sin \theta \,\sin\phi +
u_3 \cos \theta \biggr]^J
\ee

\noindent for some normalization ${\cal N}_J(u)$. If we choose
$u_1=u_2=0$ and $u_3 = \pm i u_4 = 1$, i.e. the operator
$\Tr(\Phi_3 \pm i \Phi_4)^J/\sqrt{\l^J J}$, then ${\cal N}_J(u) =2^{-J/2}$.

\section{R-symmetry }\label{app3}

Let $ {\cal O}_J = \frac{1}{\sqrt{J\,\l^J}}\,\Tr
\left(\Phi_1 + i \Phi_2\right)^J $,
   Let $U$ be a rotation in
the $x^1$-$x^2$ plane. Then

\be \la {\cal O}_J(x) \, W[C_{\tiny 1/4}] \ra  = \la U\, {\cal
O}_J(x) \, W[C_{\tiny 1/4}]  \, U^\dag \ra = \la {\cal O}_J(U\,x) \,
U \, W[C_{\tiny 1/4}]  \, U^\dag \ra \ee  Examining $C_{\tiny 1/4}$
in (\ref{quarterloop}), we see that the spatial rotation acting on
$W[C_{\tiny 1/4}] $ may be realized by a shift in the contour
parameter $\t$, which can in turn by compensated by an R-symmetry
rotation $R$ in the $\T^1$-$\T^2$ plane, $ U\, W[C_{\tiny 1/4}]  \,
U^\dag = R\, W[C_{\tiny 1/4}]  \, R^\dag $.  Then,

\be \la {\cal O}_J(x) \, W[C_{\tiny 1/4}]  \ra = \la R\, {\cal
O}_J(Ux)\, R^\dag\, W[C_{\tiny 1/4}] \ra.  \label{proof}\ee The
operator expansion coefficient depends on the leading asymptotic in
large $x$ which is a function of only the length of $C_{\tiny 1/4}$
and $x^2$,

\be \la {\cal O}_J(x) \, W[C_{\tiny 1/4}]  \ra \simeq
\left(\frac{2\pi R}{4\pi^2 x^{2}}\right)^J\xi_J+\ldots
\label{proof1}\ee Performing the $\T^1$-$\T^2$ plane R-symmetry
transformation on ${\cal O}_J$ multiplies it by a phase $\exp(i J
\phi)$ so that,

\be \la R\, {\cal O}_J(Ux)\, R^\dag\, W[C_{\tiny 1/4}]  \ra \simeq
e^{iJ\phi}\left(\frac{2\pi R}{4\pi^2 (Ux)^{2}}\right)^J\xi_J+\ldots
= e^{iJ\phi}\left(\frac{2\pi R}{4\pi^2 x^{2}}\right)^J\xi_J+\ldots
\ee Using (\ref{proof}) and (\ref{proof1}), we have
$e^{iJ\phi}\,\xi_J = \xi_J$, i.e. $\xi_J=0$.

\end{document}